\pdfoutput=1

\documentclass[11pt]{article}

\usepackage[final]{acl}

\usepackage{times}
\usepackage{latexsym}

\usepackage[T1]{fontenc}

\usepackage[utf8]{inputenc}

\usepackage{microtype}

\usepackage{inconsolata}

\usepackage{graphicx}
\usepackage{epigraph}
\usepackage{enumitem}
\usepackage[colorinlistoftodos, textwidth=65mm, shadow]{todonotes}

\title{Advancing Social Intelligence in AI Agents:\protect\\Technical Challenges and Open Questions}

\author{Leena Mathur\textsuperscript{1}, Paul Pu Liang\textsuperscript{2}, Louis-Philippe Morency\textsuperscript{1}\\
       Language Technologies Institute\textsuperscript{1}, Machine Learning Department\textsuperscript{2}\\School of Computer Science, Carnegie Mellon University\\
       {\texttt{\{lmathur,pliang,morency\}@cs.cmu.edu}}}

\begin{document}
\maketitle
\begin{abstract}
Building socially-intelligent AI agents (Social-AI) is a multidisciplinary, multimodal research goal that involves creating agents that can sense, perceive, reason about, learn from, and respond to affect, behavior, and cognition of other agents (human or artificial). Progress towards Social-AI has accelerated in the past decade across several computing communities, including natural language processing, machine learning, robotics, human-machine interaction, computer vision, and speech. Natural language processing, in particular, has been prominent in Social-AI research, as language plays a key role in constructing the social world. In this position paper, we identify a set of underlying technical challenges and open questions for researchers across computing communities to advance Social-AI. We anchor our discussion in the context of social intelligence concepts and prior progress in Social-AI research. 
\end{abstract}

\section{Introduction}
\label{sec:intro}
\begin{figure*}[!htb]
    \centering
\includegraphics[width=0.81\linewidth]{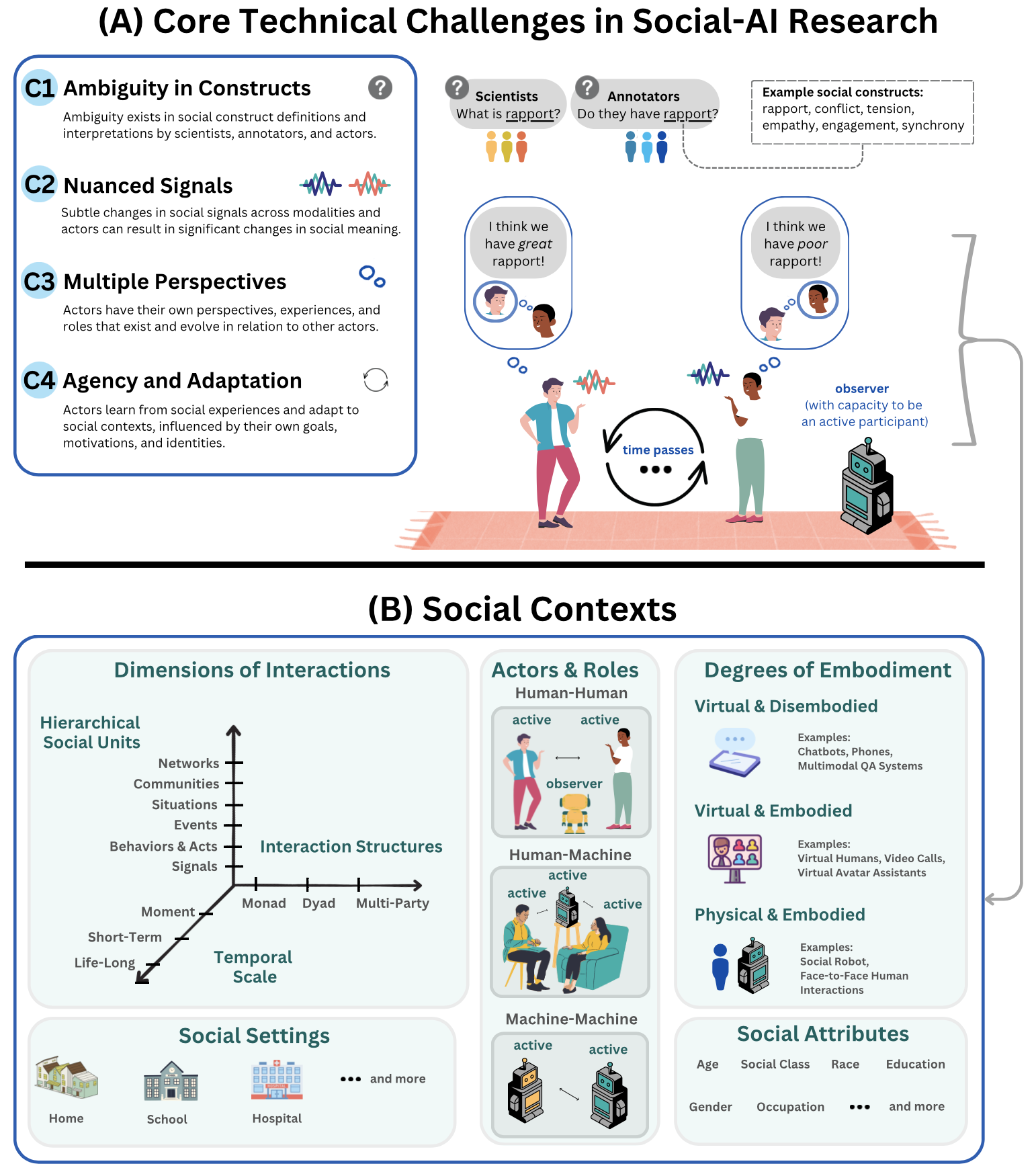}
    \caption{(A) Four core technical challenges in Social-AI research, illustrated in an example context of a Social-AI agent observing and learning from a human-human interaction. (B) Social contexts in which Social-AI agents can be situated, with interactions spanning social units, interaction structures, and timescales. Interactions can span social settings, degrees of agent embodiment, and social attributes of humans, with agents in several roles.}
    \raggedbottom
    \label{fig:complexities}
\end{figure*}

Humans rely on \textit{social intelligence} to interpret and respond to social phenomena such as empathy, rapport, collaboration, and group dynamics. Social intelligence competencies that evolved over thousands of years in \textit{Homo sapiens} are hypothesized to have been  core factors shaping human cognition and driving the emergence of language, culture, and societies \cite{wilson2012social,emery2007introduction,knight2000evolutionary, goody1995social, sterelny2007social}.  Humans today continually navigate diverse social contexts, from short-term  dyadic conversations to long-term relationships. Virtual and embodied AI agents must have social intelligence competencies in order to function seamlessly alongside humans and other AI agents. The complexity of this vision and a set of core technical challenges for developing these agents are visualized in Figure \ref{fig:complexities}.

Building socially-intelligent AI agents (Social-AI) involves developing computational foundations for agents that can sense, perceive, reason about, learn from, and respond to affective, behavioral, and cognitive constructs of other agents (human or artificial). Social-AI research interest has accelerated across computing communities in recent years, including natural language processing (NLP), machine learning (ML), robotics, human-machine interaction, computer vision, and speech (Figure \ref{fig:growth}). We see Social-AI beginning to support humans in real-world contexts. Virtual text agents have stimulated empathic conversations between humans in online chatrooms \cite{sharma2023human}, and affective signals from wearables have supported well-being \cite{sano2016measuring, park2020wellbeat}. Embodied social robots have supported geriatric care \cite{gonzalez2021social, fleming2003caregiver}, motivated stroke patients \cite{mataric2007socially, feingold2020social}, assisted youth with autism spectrum condition \cite{hurst2020social, scassellati2012robots}, improved student mental health \cite{jeong2020robotic}, and collaborated with humans in manufacturing \cite{sauppe2015social}, among other prosocial applications\footnote{While Social-AI has prosocial applications, it must be advanced in ethical ways. We discuss Social-AI ethical considerations in Section \ref{sec:ethics} and Appendix \ref{sec:infra}.}.

\raggedbottom

Is Social-AI purely an application of AI to social contexts or do underlying technical challenges emerge that are particularly relevant to Social-AI? We believe  there are core technical challenges that must be addressed to advance the multidisciplinary, multimodal  goal of building Social-AI. Our position paper is driven by the following question: 
\textit{What are core technical challenges, open questions, and opportunities for researchers to advance Social-AI?} We anchor our position in the context of social intelligence concepts, reviewed in Section \ref{sec:foundation}, and progress in Social-AI, summarized in Section \ref{sec:current_state}. In Section \ref{sec:core_challenges}, we present 4 core technical challenges, along with opportunities and open questions to advance Social-AI research.

\section{Social Intelligence Concepts}
\label{sec:foundation}
A shared understanding of social intelligence concepts is  useful for researchers to contextualize progress and challenges in Social-AI. We begin with the concept of \textit{social constructs} -- what makes entities social? Social ontologists  distinguish between \textit{social constructs} (entities that exist by human construction) and \textit{natural kinds} (entities that exist regardless of the interpretations of human minds) \cite{searle1995construction, searle1998social, searle2010making, khalidi2015three}. For example, a person is a natural kind with physical properties, but a \textit{friend} or \textit{stranger} are social constructs unlinked to those physical properties. Language plays a key role in forming social constructs (e.g., referring to someone as a \textit{close friend} is an act that can make them a close friend) \cite{searle2012human}. Social constructs are \textit{ontologically subjective}, as their existence depends on perceptions of humans, referred to as being "perceiver-dependent" \cite{searle1998social}.  This ontological subjectivity of constructs informs the Social-AI challenges that we present in Figure \ref{fig:complexities}A and discuss in Section \ref{sec:core_challenges}.  

\subsection{Social Intelligence}
\label{sec:definitions}
The term \textit{social intelligence} was introduced in the early 20th century by social scientists who recognized the importance of teaching children how to comprehend social situations \cite{dewey1909moral, lull1911moral}. Hypotheses that social intelligence differs from other forms of intelligence began in the 1920s, with distinctions among abstract intelligence (idea-focused), mechanical intelligence (object-focused), and social intelligence (people-focused) \cite{thorndike1920intelligence, thorndike1937evaluation}, with social intelligence seen as context-dependent \cite{strang1930measures}. Studying properties of social intelligence is an ongoing research area \cite{ conzelmann2013new, brown2019social}.

Humans generate and interpret social signals through verbal and nonverbal sensory channels (e.g., word choice, gaze, gesture) \cite{vinciarelli2011bridging, vinciarelli2018multimodal, burgoon2011nonverbal}.
Humans perceive social meaning from these signals \cite{poggi2012social}. A prevalent model developed with construct validation \cite{phdthesis, weis2005social, conzelmann2013new} proposes that social intelligence encompasses 5 competencies enabling humans to navigate social situations: \textit{social perception}, \textit{social knowledge}, \textit{social memory}, \textit{social reasoning}, and \textit{social creativity}. We consider a 6th competency, \textit{social interaction}. Social-AI research attempts to endow agents with these 6 competencies as key elements of social intelligence, defined below. 

\textit{Social perception} involves the ability to perceive socially-relevant information from sensory stimuli \cite{zebrowitz1990social, adolphs2016data} (e.g., reading tension from body language). \textit{Social knowledge} includes both \textit{declarative} (factual) and \textit{procedural} (norms) knowledge \cite{ snyder1980thinking, bye1993proposed}.  \textit{Social memory} involves the ability to store and recall social knowledge about the self and others \cite{skowronski1991social, nelson2003self}. \textit{Social reasoning} involves the ability to interpret social stimuli and make inferences based on these stimuli and commonsense understanding \cite{gagnon2021reasoning, read2013constraint}. \textit{Social creativity}, sometimes referred to as "theory-of-mind", involves the ability to counterfactually reason about social situations \cite{hughes2015social, astington1995theory}. \textit{Social interaction} involves the ability to engage with other agents in mutually co-regulated patterns \cite{turner1988theory, mccall2003interaction, de2010can}. 

 \subsection{Dimensions of Social Context}
 \label{sec:social_context}
Social contexts in which AI agents can be studied and deployed are visualized in Figure \ref{fig:complexities}B. These encompass diverse \textbf{social settings} (e.g., homes, hospitals) that influence social norms (e.g., silence in a library, yelling at a hockey match) \cite{rachlinski1998limits, 3f8216d3-bd87-39fc-85ee-5d4b8b511269}. Within  social contexts, \textbf{actors} (human and machine) can have different \textbf{roles} (e.g., active participant, observer) and different \textbf{social attributes} (e.g, age, occupation) which can shape interactions \cite{van2013social, trepte2013social, allen2023difference, goffman2016presentation}. The degree of \textbf{embodiment} of actors can enrich and augment communication channels \cite{goodwin2000action, wainer2006role, deng2019embodiment}; embodiment spans disembodied virtual agents (e.g., chatbot), embodied virtual agents (e.g., avatar), and physically-embodied agents (e.g., humans and robots). 

Within social contexts, there are several \textbf{dimensions of interactions}: hierarchical social units, interaction structures, and temporal scale. Social units influence communication content, norms, and interpretation \cite{hymes1972models, angelelli2000interpretation, agha2006language, goody1995social, goffman2002presentation}. To visualize social units, consider the following scenario adapted from \citet{hymes1972models}: two people exchange eye contact (\textit{social signals}), while one of them tells a joke (\textit{act}) during a conversation (\textit{event}) at a party (\textit{situation}) governed by shared norms (\textit{community}). The dyad's interaction may be viewed as two connected nodes in a \textit{social network} of interactions. Monads, dyads, and multi-party interaction structures can induce their own social dynamics \cite{pickering2021understanding}. For example, when a third actor is added to a dyad, new dynamics of group coordination and conflict can arise \cite{olekalns2003phases}. Social interactions also have a temporal scale that spans split-second communication, moments, short-term interactions, longer-term interactions, and life-long relationships \cite{wittmann2011moments}. This temporal dimension can influence social meaning conveyed during interactions; for example, a 100 millisecond pause longer than normal for an actor can indicate reluctance, instead of eagerness  \cite{durantin2017social}.

This paper proposes core technical challenges that are relevant to Social-AI agents situated across many possible dimensions of social context. 

\section{Progress in Social-AI Research} 
\label{sec:current_state}

\begin{figure}[t]
    \centering
    \includegraphics[width=0.98\linewidth]{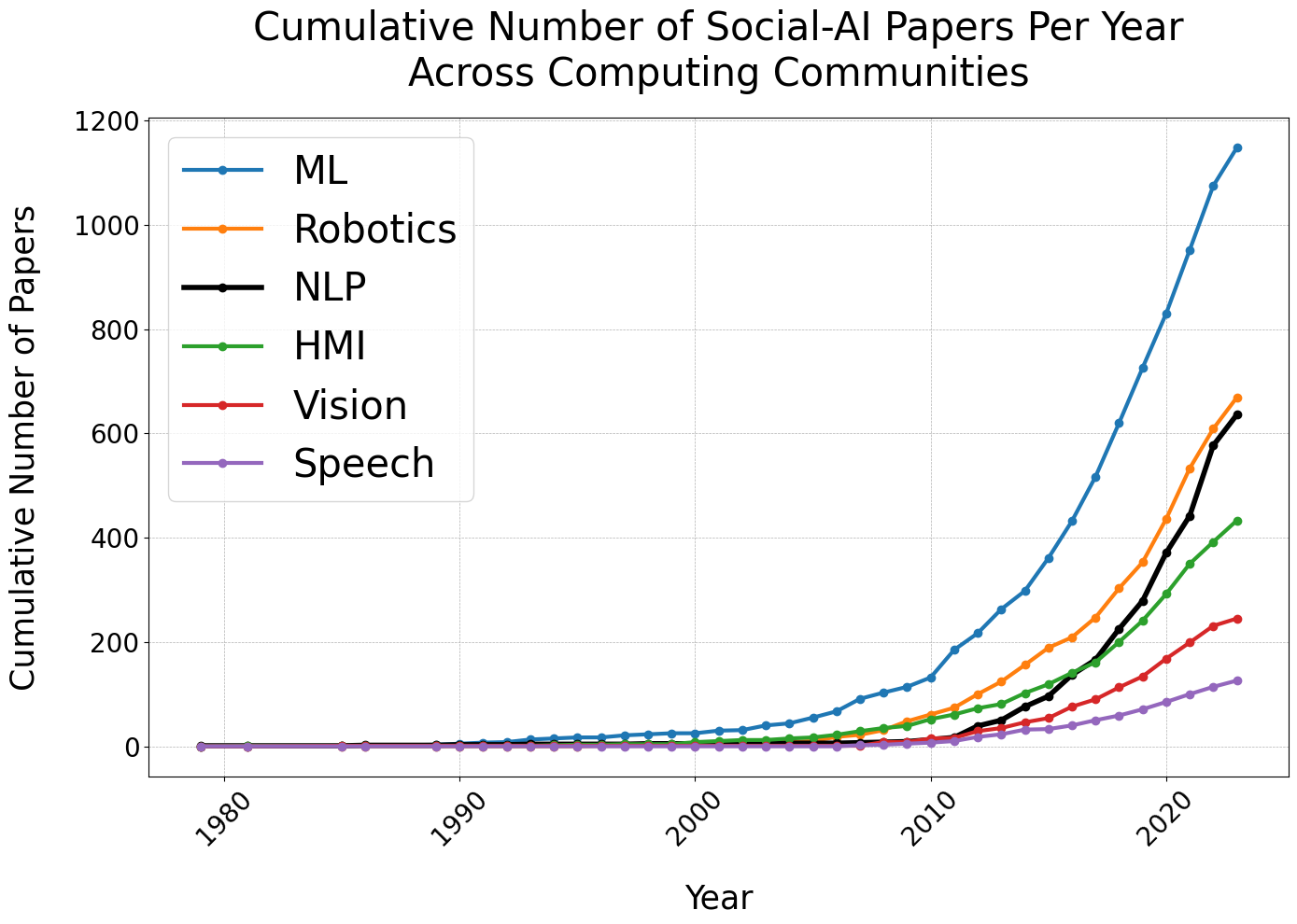}
    \caption{Cumulative number of Social-AI papers over time, based on the 3,257 papers from our Semantic Scholar Social-AI queries. Interest in Social-AI research has  been accelerating across computing communities.}
    \raggedbottom
    \label{fig:growth}
\end{figure}

We examined Social-AI research progress in 6 computing communities: NLP, ML, robotics, human-machine interaction\footnote{We use "human-machine interaction" to include both human-computer interaction and human-robot interaction.}, computer vision, and speech. Interest in Social-AI research  has accelerated in the past decade, notably in NLP, ML, and robotics (Figure \ref{fig:growth}). We synthesize key trends  to provide readers with context for the discussion of core technical challenges and open questions in Section \ref{sec:core_challenges}. Using the Semantic Scholar API \cite{kinney2023semantic}, we found 3,257 relevant papers that spanned 1979-2023. For insights, we examined representative papers across communities, selected for citation count, recency, and relevance. Details on search queries and filtering criteria are in Appendix  \ref{sec:paper_curation}.

Early Social-AI research primarily envisioned \textit{rule-based} approaches for building social intelligence competencies in agents. NLP and human-machine interaction researchers examined rule-based approaches for modeling goal-oriented communication \cite{pershina1986elementary}, cooperation \cite{d1997cooperation}, and dialogue \cite{mcroy1993abductive}. They also explored rule-based approaches for processing multimodal signals during human-machine interactions \cite{nagao1994social},
animating embodied virtual humans as conversational agents \cite{pelachaud1991linguistic, cassell1994animated, pelachaud1996generating}, and interpreting potential effects of speech acts \cite{pautler1998computational}. ML and robotics researchers proposed multi-agent search and planning algorithms for social learning in groups of disembodied agents \cite{ephrati1993multi} and \textit{behavior-based} control for groups of robots \cite{mataric1993designing, mataric1994learning}. Social robotics  advanced systems for attending to social stimuli (e.g., faces) and communicating intent \cite{812787,breazeal1999context}.

In the past two decades, scientists have increasingly leveraged \textit{ML} and \textit{deep learning} in Social-AI research. A common approach in these methods is to train models to predict social phenomena from observable human behavior, by using \textit{static} datasets with ground truth labels computed as aggregations of annotator perspectives. A focus on ML for \textit{social signal processing} also emerged during this era \cite{vinciarelli2009social}. This stimulated a focus on predicting \textit{social signals}, such as laughter from speech \cite{brueckner2014social, eyben2015geneva}, gesture and gait from visual cues \cite{morency2007latent, chao2019gaitset}, and engagement from visual, speech, and physiological data during human-robot interactions \cite{rudovic2018personalized}. There has been substantial research effort in predicting \textit{affective information}, such as emotion and sentiment, from multimodal conversation signals \cite{busso2008iemocap, schuller2012avec, morency2011towards, zadeh2018multimodal, majumder2019dialoguernn}, as well as efforts to predict social behaviors with affective information \cite{mathur2023expanding}. Multimodal interaction and computer vision research explored approaches for rendering \textit{virtual humans} with social behavior \cite{swartout2006toward, ng2022learning} and learning representations to model social interactions \cite{soleymani2019multimodal,lee2024modeling}. Computer vision and robotics researchers have studied ML approaches for predicting human \textit{intent} to inform human-robot interactions \cite{strabala2012learning} and \textit{social navigation} to improve robot navigation in spaces with humans \cite{pellegrini2010improving, kosaraju2019social, cuan2022gesture2path, taylor2022observer}. 

\raggedbottom

In parallel, researchers have focused on integrating \textit{game-theoretic} and \textit{probabilistic} approaches to model aspects of social intelligence, such as cooperation, competition, and theory-of-mind \cite{castelfranchi1998modelling, kleiman2016coordinate}. Approaches include stochastic games, inverse planning, and multi-agent reinforcement learning (MARL) \cite{jaques2019social, shum2019theory,wu2020too},  studied through formalized games, such as public goods games \cite{wang2023emergence}, or grid-world simulations \cite{lee2021joint}. 

In recent years, there have been efforts to  probe the extent to which models, in particular large language models (LLMs), exhibit social intelligence competencies. Scientists have identified strong performance by LLMs in procedurally-generated multi-agent gridworld interaction tasks \cite{kovavc2023socialai}, as well as limitations in LLM \textit{social knowledge} and \textit{social reasoning} competencies \cite{sap2022neural, shapira2023clever} when tested on \textit{static} text benchmarks such as \textsc{SocialIQa} \cite{sap2019socialiqa}, ToMI \cite{le2019revisiting}, and \textsc{Epistemic Reasoning} \cite{cohen2021exploring}. These model limitations also emerge in \textit{static} VideoQA benchmarks such as \textsc{Social-IQ 1.0} \cite{zadeh2019social}, \textsc{Social-IQ 2.0} \cite{wilf2023social}, and synthetic video benchmarks such as \textsc{MMToM-QA} \cite{jin2024mmtom}. There have been efforts to study LLM \textit{social interaction} competency during goal-oriented dialogue \cite{hosseini2020simple} and open-domain dialogue  \cite{nedelchev2020language}. Recently, \textit{dynamic} environments have been proposed to study Social-AI agents in dyadic and multi-party settings, with \textsc{Sotopia} \cite{zhou2023sotopia}, \textsc{CAMEL} \cite{li2023camel}, and \textsc{AgentVerse} \cite{chen2023agentverse} studying text agent  interactions, \textsc{CoELA} \cite{zhang2023building} studying embodied language agent interactions, and \textsc{Habitat 3.0} \cite{puig2023habitat} studying simulations of human-robot interactions. 

\paragraph{Key Takeaways} Social-AI research has made substantial progress in modeling social phenomena in \textit{static} ungrounded data (abstracting away physical and social context), synthetic data, and lab-controlled social interactions. However, these data can abstract away the \textit{richness} of multimodality \cite{liang2024quantifying} in interactions, as well as the \textit{context-sensitivity} and \textit{ambiguity} inherent to social phenomena in-the-wild. We also found a focus on modeling \textit{temporally-localized} phenomena (e.g., split-second communication, short-term interactions); longer-term phenomena were comparatively understudied (e.g., hours, days, years).

\section{Core Technical Challenges\hspace{10em}and Open Questions}
\label{sec:core_challenges}
Social intelligence concepts (Section \ref{sec:foundation}) and prior research in Social-AI\footnote{We contribute a repository of resources to inform researchers addressing challenges, discussed in Appendix \ref{sec:infra} and linked here: \url{https://github.com/l-mathur/social-ai}} (Section \ref{sec:current_state}) informed our identification of 4 core technical challenges visualized in Figure \ref{fig:complexities}A:  \textbf{(C1)
 ambiguity in constructs, (C2) nuanced signals, (C3) multiple perspectives,  (C4) agency and adaptation}. We believe these challenges are particularly relevant to Social-AI and must be addressed to advance social intelligence in AI agents situated across the landscape of social contexts in Figure \ref{fig:complexities}B\footnote{We sought to disentangle \textit{scale} (e.g., number of actors and constructs) from the identification of core  challenges.}. In this section, we present these challenges, along with opportunities and open questions for Social-AI research. 

\subsection{(C1) Ambiguity in Constructs}
\label{sec:ambiguity}
\textit{Social constructs have inherent ambiguity in their definition and interpretation in the social world}. For example, consider the  interaction in Figure \ref{fig:complexities}A -- how might we characterize rapport, conflict, or tension between the actors? Many of these social constructs are still being defined and operationalized by scientists \cite{neequaye2023rapport, policarpo2015friend}. The ontological subjectivity of social constructs (entities that exist by human construction, as discussed in Section \ref{sec:foundation}) results in inherent \textbf{ambiguity in construct definitions}. This ambiguity is amplified when defining and measuring \textbf{hierarchical constructs} -- social constructs composed of other social constructs. For example, 
it has been theorized that rapport between humans can be measured by composing estimates of mutual attentiveness, positivity, and coordination \cite{tickle1990nature}.  How might we interpret and measure these inherently ambiguous components? This is compounded when quantifying observations of a social construct ("some rapport", "a little conflict", "a lot of tension").
When modeling ambiguous social constructs, there is likely to also be misalignment in interpretations of these constructs by actors within interactions and annotators viewing interactions. This misalignment amplifies \textbf{ambiguity in ground truth} of social constructs. For example, in Figure \ref{fig:complexities}A, it is challenging to assign a label that conclusively represents rapport in the interaction. 

\paragraph{C1 Opportunities and Open Questions}
Researchers must reconsider methods for representing ambiguous social constructs in modeling approaches. When there exists ambiguity in social construct ground truth, as observed in annotator ratings \cite{yannakakis2015ratings}, how might we incorporate this ambiguity in Social-AI modeling approaches? 
Modeling efforts have largely relied on predefined label sets and "gold standard" annotations that aggregate annotators' interpretations of social constructs into discrete labels (computed by majority vote) or continuous labels (computed by temporal alignment) \cite{nicolaou2014dynamic, kossaifi2019sewa}. However, annotators' and actors' definitions and interpretations of ambiguous social constructs can vary during the course of interactions. We believe such social constructs cannot be solely represented with a single discrete or continuous number on a \textit{static} scale for an interaction. 

Natural language, with its ability to expressively represent concepts \cite{liu2004commonsense} and construct the social world \cite{searle1995construction}, may offer a uniquely effective modality to help address sources of ambiguity when representing and modeling social constructs. We see considerable opportunity for  Social-AI researchers to explore techniques for leveraging richer natural language descriptions of social phenomena in label spaces. Researchers can explore approaches for constructing flexible natural language label spaces that are \textit{dynamically-generated} and \textit{adjusted} during training and inference, instead of relying on predefined static labels to categorize inputs. These flexible label spaces would enable the development of models for social constructs that can represent greater ambiguity in social contexts. How to best design frameworks to accommodate flexible and dynamic label spaces when modeling social constructs remains an open question and research opportunity. 
 
We direct readers to representative work in areas that can inform Social-AI research towards this challenge: questioning the assumption that labels in ML tasks must have one "ground truth" \cite{aroyo2015truth, cabitza2023toward}, exploring ordinal representations (e.g., relative scales) for emotion annotation \cite{yannakakis2017ordinal}, modeling perception uncertainty across annotators \cite{han2017hard, ghandeharioun2019characterizing}, inferring annotator subjectivity across samples \cite{sampath2022seedbert}, evaluating the alignment of "ground truth" annotations with diverse annotator perspectives \cite{santy2023nlpositionality}, and handling varied label distributions (e.g., label distribution learning, multi-label learning) \cite{geng2016label, liu2021emerging}. 


\subsection{(C2) Nuanced Signals}
\textit{Social constructs are expressed through behaviors and signals that can be nuanced, often manifesting through different degrees of synchrony across actors and modalities}. During interactions, small changes in social signals  can lead to large shifts in social meaning being conveyed. For example, an actor's slight change in posture or split-second vocal emphasis on a particular word  can communicate rapport. The challenge lies in enabling Social-AI agents to \textbf{perceive and generate fine-grained social signals} (e.g., chatbot sensing a user's slowly-building frustration, robot making subtle gestures). Social signals can be expressed with different degrees of \textbf{synchrony} and can be \textbf{interleaved} across actors and across modalities, with actors functioning as both speakers and listeners \cite{morency2010modeling, watzlawick2011pragmatics}. The challenge involves advancing social signal processing techniques \cite{shmueli2014sensing} and multimodal models that operate upon verbal and nonverbal information to interpret the nuances of cross-actor and cross-modal interaction patterns.


\paragraph{C2 Opportunities and Open Questions:}
While scientists are studying the capacity of language to scaffold visual understanding \cite{el2023learning,rozenberszki2022language}, audio understanding \cite{elizalde2023clap}, and virtual agent motion generation \cite{zhai2023language, zhang2023motiongpt}, the role of natural language supervision for processing nuanced multimodal social signals remains an open question. We anticipate that advancing the ability of Social-AI systems to process highly-nuanced signals will require researchers to critically examine the role of language in guiding social perception and behavior generation. To what extent can language be treated as an \textit{intermediate representation} \cite{zeng2022socratic} to represent and integrate nuanced multimodal social signals? Are there social signals during interactions that cannot be effectively described in language? We expect that processing fine-grained signals such as vocal cues, eye movements \cite{adams2017social}, and gestures (often interleaved within milliseconds), alongside natural language, will require Social-AI researchers to develop new frameworks for representing and aligning social signals across multiple actors and multiple modalities. 

Another modeling consideration related to this challenge lies in developing mechanisms for agents to perceive the \textit{absence} of cues. Most ML models are trained to learn representations of attributes based on the \textit{presence} of those attributes in data, yet much nuance in social perception lies in recognizing the absence of stimuli, such as unsaid words, omitted eye contact, silences, and failure to adhere to social norms. Humans are theorized to learn from \textit{absent} cues in addition to visible cues, per error-driven and implicit learning frameworks \cite{markman1989lms, van1994cue, nixon2022does}. Researchers have considerable opportunity to explore approaches for addressing the \textit{absence} of cues in algorithms for agent perception of nuanced social signals. 

It remains unknown how abilities in nuanced social perception or social behavior generation might emerge in models during training or fine-tuning \cite{gururangan2020don, wei2022emergent, arora2023theory}, leading to the following questions: What are the capabilities and limitations of training objectives, such as next-token prediction or masking, in inducing nuanced social understanding in models? How might the type and amount of training data influence a model's abilities in social understanding? How might tokenization schemes influence a model's abilities in social understanding? While inductive biases of tokenization \cite{singh2024tokenization} and fine-grained understanding in models have been studied in non-social domains (e.g., segmentation, spatial relationships) \cite{krause2015fine, bugliarello2023measuring, guo2024regiongpt}, fine-grained social  understanding abilities have been understudied. Research in this direction would advance community understanding of how social intelligence competencies might (or might not) emerge in models under various training paradigms. Datasets with nuanced social signals will be essential to pursue these open questions; we discuss data considerations in Appendix \ref{sec:infra}.

\subsection{(C3) Multiple Perspectives}
\textit{In social interactions, actors bring their own perspectives, experiences, and roles; these factors can change over time and influence the perspectives of other actors during interactions.} The \textbf{subjectivity} in each actor's perceptions of social situations stems from the "perceiver-dependent" nature of social constructs (Section \ref{sec:foundation}). For example, in the interaction in Figure \ref{fig:complexities}A, both actors have different perspectives on the level of "rapport" in their interaction. The evolution of these \textbf{concurrent multiple perspectives} can be influenced by several factors of the interaction's social context. Actors meeting for the first time will use different information to estimate rapport than if they had a long-term relationship with frequent interactions to draw upon. The topic of conversation, social setting (e.g., are they in a doctor's office?), behavioral norms \cite{ziems2023normbank}, social roles (e.g., are they colleagues?) and additional social attributes (e.g., age) can influence the evolution of their concurrent perspectives. 

In addition, there exists \textbf{multi-perspective interdependence}, as each actor's perspective can change over time, while influencing and being influenced by others' perspectives. For example, a hospital patient and assistive Social-AI agent interacting intermittently would be continually adjusting their perspective of the other actor to build rapport over time. The challenge involves equipping Social-AI agents with the capacity to  reason over these multi-perspective dynamics. 

\paragraph{C3 Opportunities and Open Questions:} 
Addressing this challenge will require modeling paradigms that enable agents to reason over multiple, dynamically-changing perspectives, experiences, and roles in social interactions. Each actor can influence and be influenced by other actors, as formalized in interactionist theories from  psychology such as  \textit{actor-partner} frameworks \cite{kenny2010detecting}, \textit{social identity theory} \cite{hogg2016social}, and \textit{social influence theory} \cite{friedkin2011social}. This complexity leads us to identify the following open questions: How can researchers create models for Social-AI agents to perceive concurrent, interdependent perspectives of actors during interactions? To what extent would a single, joint model be more effective than multiple models (e.g., one for each actor) to represent social phenomena across an interaction? When interactions occur intermittently over time, how can models efficiently and accurately adjust agents' perceptions of other actors' perspectives? 

The social intelligence competencies of social creativity (theory-of-mind) and social reasoning (Section \ref{sec:definitions}) are typically associated with this "multi-perspective" modeling challenge. Existing theory-of-mind research in Social-AI has focused on the movement of abstract shapes \cite{gordon2016commonsense}, procedurally-generated dialogues \cite{kim2023fantom}, and synthetic videos \cite{jin2024mmtom} and has been useful in assessing abilities of models in these tasks. However, we believe that the multi-perspective Social-AI challenge encompasses not only the capacity to reason about the mental states and beliefs of other agents (including higher-order theory-of-mind), but also the capacity to perform counterfactual social reasoning \textit{over time} with multiple dimensions of dynamic social contexts  influencing interdependent perspectives of the actors. In order to test models' ability to address this challenge, the Social-AI community will require the curation of fine-grained benchmarks that test multi-perspective abilities in naturalistic social settings. Data  considerations for the Social-AI community are discussed in Appendix \ref{sec:infra}. 

\subsection{(C4) Agency and Adaptation}
\label{sec:agency}
\textit{Actors learn from social experiences and adapt to social contexts, through interactions, influenced by their own agency, goals, motivations, and identities.} Social-AI agents must have the capacity to be \textit{goal-oriented}, often targeting multiple goals simultaneously. For goal-oriented Social-AI agents to \textbf{learn from social experiences}, these agents must have mechanisms for motivation to \textbf{adapt to both explicit and implicit social signals} from other actors and from the surrounding social context. \textit{Explicit social signals} can include direct or rank-based feedback provided by humans assessing the perceived competencies of an AI agent driven by a social goal. \textit{Implicit signals} can include verbal and nonverbal cues from actors, as well as latent aspects of the social environment (e.g., community-level norms) that can provide supervision for a goal-oriented agent. For example, in Figure \ref{fig:complexities}A, given a social goal of faciliting rapport, the robot might use its observations of the dyad's explicit and implicit social signals to learn how to successfully intervene in the interaction. Creating goal-oriented Social-AI agents will involve developing computational mechanisms for \textbf{shared social memory} between a Social-AI agent attempting to learn from social experiences and other actors in an interaction. Building a shared social reality across actors within an interaction \cite{echterhoff2009shared} enables agents to create common ground \cite{clark1991grounding} and operate in ways aligned with mutual social expectations. Alignment in these social expectations will be necessary for Social-AI agents to effectively adapt behavior in relation to other actors and achieve long-term social goals. 

\paragraph{C4 Opportunities and Open Questions:} 
For many non-social tasks, motivation for agents to learn how to exhibit certain behaviors can be instilled through clearly-defined loss functions and metrics (e.g., minimizing bounding box error for object detection), as well as reward signals based on these metrics. However, Social-AI agents need to learn from multiple kinds of \textit{social experiences} \cite{hu2022towards}. In many cases, this can involve learning from implicit social signals that are \textit{fleeting} (e.g., vocal cue lasting 1 second), \textit{sparse} (e.g., an actor raising an eyebrow once to indicate disapproval), and \textit{context-dependent} (e.g., varied cultural norms). Learning from these types of social signals will be important, as it is infeasible to expect humans to provide regular, explicit feedback to indicate satisfaction with a Social-AI agent's competencies. It is possible that humans will be unaware of the Social-AI agent's goals, reducing the likelihood of humans providing explicit feedback in-the-wild about the agent's performance. 

The challenge for researchers is to build mechanisms that motivate Social-AI agents to learn from diverse spaces of social signals. This challenge leads us to identify several understudied, open questions: How can modeling paradigms and metrics be created for Social-AI agents to estimate success in achieving social goals, based on explicit and implicit signals? How can mechanisms to adapt behavior towards achieving \textit{single goals} and  \textit{multiple, simultaneous goals} be developed for agents? How might social rewards \cite{tamir2018social} shape agent learning over time? How can shared social memory be built among Social-AI agents and other actors in interactions, and how can this memory inform algorithms for learning from social signals? 
To begin addressing these directions, researchers might consider perspectives on motivation \cite{chentanez2004intrinsically}, learning from implicit signals  \cite{jaques2020human, wang2022affective}, value internalization \cite{rong2024value} from social feedback, and agent memory \cite{cheng2024exploring, zhong2024memorybank}. There exists considerable opportunity to tackle open questions related to Social-AI agency and adaptation.

\section{Conclusion}
\label{sec:conclusion}
In this position paper, we present a set of core technical challenges, along with opportunities and open questions, for advancing Social-AI research across computing communities. The core technical challenges that we identify  include the following: (C1) ambiguity in constructs, (C2) nuanced signals, (C3) multiple perspectives, and (C4) agency and adaptation in social agents. We believe these challenges are relevant for developing Social-AI agents situated in diverse social contexts. For example, NLP researchers situating chatbots within social dialogue and robotics researchers building social robots will both encounter these 4 challenges. 

Our research vision for building Social-AI encompasses work to advance social agents with a variety of embodiments, social attributes, and social roles, with agents interacting in a range of social contexts. We conceptualize dimensions of social context and  illustrate the core technical challenges in Figure \ref{fig:complexities}, as a resource for readers across fields to visualize a holistic perspective of Social-AI. 

We anchor our paper in key perspectives from the social intelligence literature (Section \ref{sec:foundation}) and progress in Social-AI research across multiple computing communities and multiple decades  (Section \ref{sec:current_state}). We contribute these summaries to stimulate a shared community understanding on Social-AI. 

Virtual and embodied AI systems with social intelligence have considerable potential to support human health and well-being in real-world contexts, such as homes, hospitals, and other shared spaces. For AI agents to function seamlessly alongside humans, these agents need to be endowed with social intelligence competencies. Our paper is intended to inform and motivate research efforts towards AI agents with social intelligence.

\section{Limitations}
\label{sec:limitations} 
\paragraph{Position Paper Constraints} This work is a \textit{position paper}, and the core challenges we identify are informed by our 
involvement in Social-AI research and our reflection on existing Social-AI papers across NLP, ML, robotics, human-machine interaction, computer vision, and speech. 
Our search queries and filtering criteria were designed to help us capture and convey trends in Social-AI research from 6 computing communities over multiple decades (Section \ref{sec:current_state} and Appendix \ref{sec:paper_curation}). We have highlighted representative papers across fields, and this position paper is not an exhaustive survey.  

\paragraph{Scope} Our paper is scoped to focus on core technical challenges and open questions in Social-AI research, anchored the social intelligence concepts and competencies discussed in Section \ref{sec:foundation}. An important  direction of research exists in \textit{social bias} detection and mitigation \cite{sap2019social, blodgett2020language, liang2021towards, lee2023survey}. Social-AI research and the deployment of Social-AI systems must be informed by bias and ethics considerations, as discussed in Section \ref{sec:ethics}. 

\section{Ethics}
\label{sec:ethics}
Social-AI has several prosocial applications to democratize human access to healthcare, education, and other domains (Section \ref{sec:intro}). Our society faces widening \textit{care gaps} with a shortage of human care providers in mental healthcare \cite{pathare2018care} and geriatric care \cite{redfoot2013aging}, alongside growing global education inequalities \cite{attewell2010growing}. Social-AI agents have potential to augment human ability to tackle these challenges; for example, through personalized Social-AI agents for education \cite{park2019model, spaulding2022lifelong, dumont2023promise}. However, while addressing all the Social-AI technical challenges, research must be conducted with ethical and privacy-preserving practices. For example, a scientist advancing the ability of AI systems to detect and generate nuanced social signals (C2) must acknowledge and mitigate the risk of undermining human \textit{trust} in AI systems that are likely to exhibit uncanny behavior in this process \cite{mori2012uncanny, mathur2016navigating}.   

\paragraph{Participatory Social-AI Research} In order to better align development and deployment of \textit{societally-beneficial} Social-AI, we advocate that future directions include embracing Participatory AI frameworks \cite{bondi2021envisioning, birhane2022power, zhang2023deliberating}, consciously involving a diverse range of stakeholders, to ensure Social-AI researchers prioritize concerns, ethical frameworks, risks, and needs raised by stakeholders. When humans interact with computers, virtual agents, and robots, they impose social norms and expectations on these agents \cite{nass2000machines}; it is important to directly assess and center human social and functional expectations from Social-AI agents when creating and deploying them \cite{takayama2008beyond, dennler2022using, olatunji2024immersive}. Participatory AI practices will help Social-AI researchers develop guidelines for \textit{transparency} \cite{felzmann2020towards} in how data is gathered, what data is stored, and how data is used through collaboration between end users of Social-AI and researchers of Social-AI. We envision that Participatory Social-AI research would involve advancing \textit{user-centric} modeling paradigms that provide users with power to choose, edit, and delete their data used in models, as well as signal \textit{social} and \textit{cultural} norms they would like their systems to follow.

\paragraph{Social Bias}
Social-AI systems often rely on models that can exhibit and amplify \textit{social bias}. Techniques to identify and mitigate social bias in language models \cite{liang2021towards, kaneko2022unmasking, prabhumoye2021few} and multimodal models \cite{luccioni2024stable, cho2023dall} are important, in order to build Social-AI systems that are culturally-competent \cite{bhatt2024extrinsic, bhatia2024local}, develop techniques that perform across cultures \cite{hershcovich2022challenges, mathur2023towards}, and refrain from propagating harmful social biases when deployed in-the-wild. 

\paragraph{Privacy and Trust}
We believe that Social-AI researchers must prioritize user \textit{privacy}, especially due to the sensitive nature of social phenomena and the deployment of Social-AI agents in spaces alongside humans \cite{kaminski2016averting}. Respecting user privacy and building \textit{trust} can involve transparency about how human data is being used, models refraining from tasks they cannot perform \cite{akter2024visreas}, and minimizing the collection and use of invasive data \cite{stapels2023never}. Models that can learn from minimal data (e.g., a single user's interaction), perform decentralized learning \cite{guerdan2022decentralized}, and store de-identified representations of  data can support the privacy of end-users. In addition, developing \textit{on-device} models capable of operating on single devices (e.g., smartphones and wearables) can minimize dependence on external services and reduce risk of exposing sensitive data. These approaches can promote \textit{trustworthiness} of Social-AI systems. 

\section*{Acknowledgements}

This material is based upon work partially supported by National Institutes of Health awards R01MH125740, R01MH132225, R01MH096951 and R21MH130767. Leena Mathur is supported by the NSF Graduate Research Fellowship Program under Grant No. DGE2140739. Paul Pu Liang was supported in part by a Siebel Scholarship and a Waibel Presidential Fellowship. Any opinions, findings, conclusions, or recommendations expressed in this material are those of the authors and do not necessarily reflect the views of the sponsors, and no official endorsement should be inferred. Figure \ref{fig:complexities} includes icon material available from https://icons8.com. 

\bibliography{custom}
\newpage
\appendix

\section{Search Queries and Filtering Criteria}
\label{sec:paper_curation}
We accessed data from Semantic Scholar \cite{lo2019s2orc,  kinney2023semantic}, a scientific research tool from the Allen Institute for AI. We bulk-downloaded meta-data of papers that (1) had ``Computer Science" as one of its fields and (2) satisfied the following Semantic Scholar query: \texttt{["social" AND "model"] OR ["social" and "agent"] OR ["social" and "artificial intelligence"] OR ["social" and "machine learning"]}. This initial search yielded approximately 161,000 papers. We, then, filtered our sample to include relevant papers from venues in the "top 20" per field by Google Scholar's venue-listings across the 6 computing communities of artificial intelligence (this Google Scholar listing includes key machine learning conferences), computer vision, natural language processing, robotics, and human-computer interaction (this Google Scholar listing includes human-machine interaction and human-robot interaction). This filtering step yielded  3,257 papers.

On the assumption that the papers between 2010 and 2020 would be the most influential in ascertaining priorities for each field and papers from 2020 onwards capture the most recent priorities of each field, we sampled 20 papers per computing community. 40\% of papers were from 2010-2019, and 60\% of papers were from 2020-2023 (15\% per year). Retaining for relevance, we pick the most cited 8 papers from the past decade and the most cited 3 papers from each year of 2020-2023. We examined  these 120 papers to help ensure that our discussion of prior work captures research priorities of computing communities in Social-AI. 

\section{Social-AI Infrastructure}
\label{sec:infra}  

We discuss several data infrastructure considerations to supplement Social-AI research efforts towards the core technical challenges in our paper. 

\paragraph{Social-AI Data Sources}
Commonly-used sources of data for Social-AI model training and evaluation are \textit{static} (e.g., lab-collected interactions, TV shows) and do not contain a naturalistic distribution of long-tail social phenomena, nuanced social signals, multi-perspective dynamics, and other dimensions of social context visualized in Figure \ref{fig:complexities}B. The majority of online video data, for example, are sociotechnical constructions \cite{knoblauch2019videography}, influenced by  choices such as frame angles and scene cuts. 

Social-AI research may benefit from emphasizing richer sources of naturalistic interactions, as well as \textit{interactive} sources of data. Examples of recent work exploring richer sources of social interaction data include multimodal egocentric perception \cite{grauman2023ego} across audio, motion/position, gaze, vision, pose, and natural language, as well as a multi-year study of multimodal neural and psychological factors related to human social cognition \cite{kliemann2022caltech}. Multimodal, ethically-collected data sources that pair natural language, vision, and other neural and sensory modalities (e.g., olfaction, physiology) during social interactions can be invaluable sources of data to advance Social-AI research. We also believe there is  great potential to explore techniques for re-purposing and augmenting existing social interaction datasets to supplement the curation of real-world datasets (e.g., reconstructing 3D humans from 2D videos) \cite{pavlakos2022one}.  

\paragraph{Social-AI Benchmarks}
Curating multimodal datasets of \textit{interactions in-the-wild} that include Social-AI agents and actors situated within multiple dimensions of social context will be useful to advancing Social-AI research. There is a scarcity of shared, longitudinal datasets of real-world interactions across social contexts that can support researchers in stress-testing algorithms to address core technical challenges in Social-AI research. For example, a researcher developing an algorithm to address the challenge of modeling multiple perspectives during social interactions might find it useful to explore how effective the approach is when tested for the same actor in different \textit{interaction structures}, spanning dyadic, multi-party, and group contexts. There is currently no shared infrastructure for systematically performing these experiments to inform our understanding of Social-AI agent performance along various dimensions of social context. This type of understanding is important to safely deploy Social-AI agents in-the-wild. 

\paragraph{Social-AI Data Annotation} Obtaining fine-grained annotations of multiple social constructs during social interactions is a resource-intensive task. In the case of social interaction video data, a fine-grained annotation might include labels at different hierarchical social units (e.g., behaviors, events) that may overlap in video segments. Each minute of an interaction can take more than an hour for a trained annotator to analyze and annotate \cite{morency2010modeling}. It is important to include multiple annotators for each sample and fairly compensate annotators for these types of cognitively-demanding annotations \cite{huang2023incorporating}. 

The challenge of ambiguity in social constructs amplifies the difficulty of this annotation process, since annotators can interpret constructs differently, even when provided with the same initial annotation instructions. It is important to design annotation paradigms that consciously include annotators from diverse backgrounds, in order to represent multiple perspectives on social constructs that are likely to exist across gender, age, culture, nationality, and other demographic factors \cite{ding2022impact, santy2023nlpositionality}.

In addition, annotators' social perception speeds can also differ when interpreting social interactions, causing \textit{temporal delays} between occurrences of social phenomena and labels across annotators. 
Techniques to address these temporal delays in annotations must be developed and standardized in modeling paradigms by Social-AI researchers. 

\paragraph{Community Resource} To accompany the paper and stimulate community research, we contribute 
a repository of resources for researchers interested in addressing Social-AI challenges: \url{https://github.com/l-mathur/social-ai}. This repository contains links to papers, books, doctoral dissertations, datasets, benchmarks, simulation environments, and university courses relevant to Social-AI research. This resource will be continually updated for the research community.

\end{document}